\documentclass[conference]{IEEEtran}
\IEEEoverridecommandlockouts
\usepackage[T1]{fontenc}
\usepackage[utf8]{inputenc}
\usepackage{cite}
\usepackage{amsmath,amssymb}
\usepackage[scaled=0.83]{beramono}
\usepackage{booktabs}
\usepackage{tabularx}
\usepackage{array}
\usepackage{listings}
\usepackage{xcolor}
\usepackage{tikz}
\usetikzlibrary{arrows.meta,positioning,calc,fit}
\usepackage[hyphens]{url}
\usepackage[hidelinks]{hyperref}
\usepackage{orcidlink}
\usepackage{microtype}
\newcommand{\code}[1]{{\fontshape{n}\ttfamily #1}}
\newif\ifdraft \draftfalse

\newcommand{\nDesigns}{50}
\newcommand{\nProved}{45}
\newcommand{\nNotCovered}{5}

\newcommand{\nCombinational}{25}
\newcommand{\nMaxPaired}{9}

\newcommand{\nMutDesignsGiven}{42}
\newcommand{\nMutDesigns}{40}
\newcommand{\nMutCompiled}{281}
\newcommand{\nMutSimDiffers}{17}
\newcommand{\nMutProofFinds}{14}
\newcommand{\nMutReaderRejects}{3}
\newcommand{\nMutAccepted}{0}
\newcommand{\nMutSimEqual}{264}
\newcommand{\nMutEqualAccepted}{255}
\newcommand{\nMutEqualRefused}{9}
\newcommand{\nMutNoCompile}{70}
\newcommand{\nMutNoCompileRead}{56}

\newcommand{\nGolden}{8}

\newcommand{\nCodeBases}{16}
\newcommand{\nModules}{2,184}
\newcommand{\nMatch}{1,185}
\newcommand{\nMatchBeforeCG}{1,127}
\newcommand{\nClockRejBefore}{65}
\newcommand{\nClockRejAfter}{13}
\newcommand{\nMismatch}{5}
\newcommand{\nRejected}{994}

\newcommand{\nPlaceholder}{557}
\newcommand{\nSteps}{3,000}

\newcommand{\nProofRecords}{1,188}
\newcommand{\nCorpusProved}{600}
\newcommand{\nCorpusBounded}{129}
\newcommand{\nCorpusFail}{45}
\newcommand{\nCorpusNotCovered}{414}
\newcommand{\nCorpusUnknown}{27}
\newcommand{\nCorpusUnsupported}{9}
\newcommand{\nCorpusNoOutputs}{6}
\newcommand{\nCorpusFalseDiff}{3}
\newcommand{\nCorpusComb}{314}
\newcommand{\nCorpusMaxRegs}{1,026}

\newcommand{\nProbes}{18}
\newcommand{\nCorpusRtl}{8}

\newcommand{\nRustLines}{25,700}

\hypersetup{
  pdftitle={Translation Validation of SystemVerilog-to-SystemC RTL Conversion by Reading the Generated C++},
  pdfauthor={Mayur Kubavat},
  pdfkeywords={SystemVerilog, SystemC, RTL, translation validation, equivalence checking}
}
\begin{document}
\title{Translation Validation of SystemVerilog-to-SystemC RTL Conversion by Reading the Generated C++\thanks{This work was done independently, on personal equipment and in a personal capacity. It is not affiliated with or endorsed by any employer.}}
\author{\IEEEauthorblockN{Mayur Kubavat\,\orcidlink{0009-0009-1087-8558}}
\IEEEauthorblockA{\textit{Independent Researcher, Mountain View, CA, USA}}}
\maketitle
\thispagestyle{plain}

\begin{abstract}
SystemC models of register-transfer level (RTL) designs are used for virtual platforms, firmware
bring-up and simulation without a commercial license, and more of them now come from translators
and language models. A wrong model usually compiles, and it usually agrees with the RTL.
SystemVerilog evaluates an 8-bit sum in 8 bits while C++ evaluates it in 64, so a translation that
forgets one truncation keeps a carry the RTL drops, and only inputs near the top of the range show
it. I built \code{svcc}, an open-source SystemVerilog-to-SystemC converter, and my first check of
its output read the generated C++ with SystemVerilog's rules. That check passed two wrong models.
The check that replaced it gives the generated header the meaning C++17 and SystemC give it and
turns undefined behavior into proof obligations. It prints what it read as SystemVerilog with every
width change explicit, then proves that equivalent to a netlist of the source by induction over
registers matched by name. On \nDesigns{} author-written designs it proves \nProved{} and reports
\nNotCovered{} as outside its model. In a mutation experiment it accepted none of the
\nMutSimDiffers{} mutants that compiled SystemC showed to be wrong. Of \nModules{} modules from
\nCodeBases{} open-source code bases, \nMatch{} conversions match Verilator in random co-simulation
and the check proves \nCorpusProved{} of them. The same runs found defects in the front end, in
\code{svcc} and in Verilator.
\end{abstract}


\section{Introduction}
\label{sec:intro}

I have worked in ASIC and SoC design verification for more than ten years, and much of that work
runs on C++ models of RTL: reference models inside scoreboards, virtual platforms that boot
firmware before silicon, and SystemC~\cite{ieee1666_2023} models that run without a simulator
license. Verilator~\cite{snyder2020verilator} turns RTL into fast C++, and with \code{-{}-sc} into a
SystemC wrapper, but its output is a simulation engine that nobody is meant to read or edit.
Readable models come from hand translation, from
translators~\cite{bombieri2010hifsuite,syed2013vhdl2sc,zheng2026bidir}, and now from language
models asked to translate between C++ and hardware languages~\cite{collini2024c2hlsc,lyu2026autoc2v}.
Some flows convert RTL to C or C++ only so that software tools can run on it, such as software
verifiers~\cite{mukherjee2016v2c} or fuzzers~\cite{gadde2024fuzzwiz}.

None of these models is checked against the RTL it came from, and their mistakes produce no
warning. SystemVerilog sizes an expression from its context. C++ promotes small integer types to
\code{int} and converts SystemC's \code{sc\_uint<W>} to a 64-bit integer. A translation of
\code{(a + b) {>}{>} 1} that does not truncate the sum keeps the carry, and \code{a = b = 0xff}
gives \code{0xff} instead of \code{0x7f}. The two models agree on most inputs. A random regression
on such a model can stay clean for a long time, because the difference needs both operands near the
top of their range.

I started \code{svcc} in my own time to convert SystemVerilog RTL to SystemC and to prove each
generated model equivalent to its source. Translation
validation~\cite{pnueli1998tv,necula2000tv,lopes2021alive2} is the obvious way to do the second
part: check each output of the translator against its input instead of proving the translator
correct. Checking C against RTL with C's own semantics is old as well. Clarke, Kroening and Yorav
did it with bounded model checking in 2003~\cite{clarke2003cverilog}, and commercial C-to-RTL
equivalence tools do it today~\cite{koelbl2009c2rtl}. So the method here is standard. My first
check still got it wrong, in a way I think is easy to repeat, and I only found out because a
different experiment ran the generated code on a real SystemC kernel.

This paper is the record of that. Section~\ref{sec:failure} describes the first check and the two
models it passed, and the two rules I took from it. Sections~\ref{sec:convert}
and~\ref{sec:gate} describe the converter and the check that replaced the first one. In
Section~\ref{sec:eval} I measure that check on \nDesigns{} designs, on a mutation experiment judged
by compiled SystemC, and on \nModules{} modules from \nCodeBases{} open-source code bases, where it
proves \nCorpusProved{} of the \nMatch{} modules that match Verilator. The same runs found three
defects in \code{svcc}, three in its front end, and two in Verilator that are now reported
upstream. \code{svcc} is released under the Apache 2.0 license on crates.io and PyPI with all
scripts and results; a companion software paper describes the tool~\cite{kubavat2026svccsx}.

\section{Where SystemVerilog and SystemC Differ}
\label{sec:diff}

Table~\ref{tab:diff} lists the rules a translator has to bridge. In each row the same expression
can give different values in the two languages.

\begin{table}[t]
\centering
\caption{Semantic differences between SystemVerilog RTL and SystemC that a translation must bridge.}
\label{tab:diff}
\footnotesize
\setlength{\tabcolsep}{3pt}
\begin{tabularx}{\columnwidth}{@{}>{\raggedright\arraybackslash}p{0.17\columnwidth} >{\raggedright\arraybackslash}X >{\raggedright\arraybackslash}X@{}}
\toprule
Topic & SystemVerilog (IEEE 1800-2023) & C++17 / SystemC \\
\midrule
Width & Context-determined: operands extend to the widest width in the expression and the assignment target (11.6) & Integer promotion to \code{int}; \code{sc\_uint<W>} converts to \code{unsigned long long}; result width set by the C++ type \\
Signedness & Any unsigned operand makes the expression unsigned (11.8) & Usual arithmetic conversions: a mixed comparison is unsigned when the unsigned type's rank is at least the signed type's, or when the signed type cannot hold all its values \\
Undefined ops & Division by zero gives X; shifts beyond the width give 0 & Signed overflow, division by zero and shift $\ge$ width are undefined behavior \\
Update order & Nonblocking updates wait in the nonblocking assignment (NBA) region until all active events of the time step, including a derived clock's change, have run & \code{sc\_signal} updates commit at the end of every delta cycle, so a process woken by a derived clock can see values already updated \\
Values & Four states (0, 1, X, Z) & Two states in \code{sc\_uint}; four in \code{sc\_lv} \\
\bottomrule
\end{tabularx}
\end{table}

A translator that gets every row right still writes C++ that takes a careful reader to check. An
engineer who knows SystemVerilog tends to read that C++ with SystemVerilog's rules. My first check
did exactly that.

\section{How the First Check Failed}
\label{sec:failure}

The first version of \code{svcc} carried RTL through a text-based path. Expressions were kept as
strings in the intermediate representation (IR). To validate a
conversion, the check took the emitted C++, re-read it as if it were SystemVerilog, and asked a
solver whether the re-read design was equivalent to the source. It reported every design in my
suite equivalent, and I took that at face value.

What changed my mind was the comparison with another translator in Section~\ref{sec:scav}. For
that comparison the generated SystemC had to compile and run on a real SystemC kernel, against
Verilator. On the \nProbes{} probes I wrote for it, \code{svcc}'s text-based path was correct on
only 3. Two of the wrong models were designs my check had already proved (Table~\ref{tab:fooled}).

\begin{table}[h]
\centering
\caption{Two models the first check reported as equivalent.}
\label{tab:fooled}
\footnotesize
\setlength{\tabcolsep}{3pt}
\begin{tabularx}{\columnwidth}{@{}l >{\raggedright\arraybackslash}X l@{}}
\toprule
Source (8-bit) & What the C++ did & First difference \\
\midrule
\code{y = (a + b) {>}{>} 1} & shifted the 64-bit sum & \code{a=b=0xff}: \code{7f} vs \code{ff} \\
\code{lt = (a - b) < c} & compared the 64-bit difference & \code{a=3e, b=c1, c=7e} \\
\bottomrule
\end{tabularx}
\end{table}

The check could not have found these. Read with SystemVerilog's rules, \code{(a + b) {>}{>} 1}
assigned to an 8-bit target is evaluated in 8 bits, exactly as in the source. The 64-bit evaluation
exists only under C++'s rules, and the check never applied them. It was making the same mistake as
an engineer reading the C++ by eye. There was a second problem behind the first. The checked side
was derived from the translator's own intermediate form, so if the lowering and the emitter shared a
misunderstanding, both sides of the comparison inherited it.

I rebuilt the check around two rules.
\begin{description}
\item[R1] The checked side starts from the emitted header text and gives it the \emph{target}
language's meaning, including the cases the target leaves undefined.
\item[R2] The reference side is built from the source text without any of the translator's code.
\end{description}
I think both rules apply to any cross-language translator, including a language model. My
implementation meets R2 only in part: the reference and the checked side both use the
\code{svlang} front end, which the translator also uses. Section~\ref{sec:assume} treats
\code{svlang} as trusted and says what that costs.

\section{The Converter}
\label{sec:convert}

\code{svcc rtl2sc} uses \code{svlang}~\cite{kubavat2026svlang}, an open-source SystemVerilog front
end I also wrote, to parse and elaborate the design. Parameters, \code{localparam}s and generate
constructs are resolved, and every expression is bound with the width and signedness rules of
IEEE~1800-2023 clauses 11.6 and 11.8. Each node then carries its width and signedness, and every
implicit conversion becomes an explicit node. \code{svcc} lowers this to one typed module per
parameter set and emits one \code{SC\_MODULE} for each.

\emph{Emitter invariant.} Each emitted C++ subexpression evaluates to exactly the value of its
typed node. Wherever C++ would compute in a wider type, the emitter truncates explicitly. The
8-bit average of Section~\ref{sec:failure} becomes
\begin{lstlisting}[language=C++]
y.write(sc_uint<8>(sc_uint<8>(a.read() + b.read()) >> 1U));
\end{lstlisting}
It is ugly, and it is correct.

\emph{Subset.} There is no standard synthesizable subset of SystemVerilog. I follow what
synthesis tools accept~\cite{sutherland2014synth} and stay inside the Accellera synthesizable
subset of SystemC~\cite{ieee1666_1_2016}. Supported constructs are continuous assignments,
\code{always\_comb}, \code{always\_ff} and \code{always\_latch}; blocking and nonblocking
assignment; \code{if}, \code{case}, \code{casez}, \code{casex}, \code{case inside}, \code{for} and
\code{while}; module and package functions; hierarchy with parameter overrides and \code{.*}
connections; bit, part and indexed selects; packed structs, enums and arrays; memories; and vectors
wider than 64 bits (\code{sc\_biguint}). Blocking assignments are read from a local copy inside the
process and written back once. Blocks that drive different bits of one signal are merged into one
process, because a SystemC signal has one writer.

\emph{Clocks.} A register may be clocked by an input, a copy of one, or a register output. A
clock-gating cell (a latch transparent while the clock is low, ANDed with the clock) is converted
to a clock enable on the root clock, and an instance whose clock port is gated gets a module variant
with an enable input. A literal translation would run the gated registers one delta cycle late
(Table~\ref{tab:diff}, update order). I added this conversion after the first stress run
(Section~\ref{sec:stress}) rejected \nClockRejBefore{} modules for clock reasons. The proof does not cover these designs; the
evidence for this conversion is co-simulation. Other derived clocks, such as clock multiplexers,
are rejected.

\emph{Fail-closed.} A construct outside the subset is an error with a source location.
\code{svcc} also rejects two things a two-state model would silently read as 0: a signal that is
read but has no driver, and a constant select outside the declared range. X and Z map to 0, which
is the model under which both the proof and the co-simulation run.

\section{The Check: Reading the Generated C++}
\label{sec:gate}

\begin{figure}[t]
\centering
\resizebox{\columnwidth}{!}{%
\begin{tikzpicture}[font=\sffamily\scriptsize,node distance=4mm and 5mm,
  box/.style={draw,rounded corners=1pt,align=center,minimum height=8mm,inner xsep=1.5mm,fill=white},
  arr/.style={-{Stealth[length=4pt]},thick},
  sh/.style={draw,dashed,rounded corners=1pt,align=center,minimum height=8mm,inner xsep=1.5mm,fill=black!4}]
\node[box] (src) {source\\.sv};
\node[box,right=of src] (ir) {typed IR};
\node[box,right=of ir] (hpp) {generated\\header};
\node[box,right=of hpp] (rd) {C++ reader\\+ UB flags};
\node[box,right=of rd] (ew) {explicit-\\width SV};
\node[sh,below=8mm of src] (n1) {svlang\\netlist};
\node[sh,below=8mm of ew] (n2) {svlang\\netlist};
\node[box,below=8mm of hpp,fill=black!10] (z3) {registers mapped\\by name, Z3 induction};
\node[sh,above=3mm of ir,minimum height=5mm] (fe) {svlang front end};
\draw[arr] (src) -- (ir);
\draw[arr] (ir) -- (hpp);
\draw[arr] (hpp) -- (rd);
\draw[arr] (rd) -- (ew);
\draw[arr] (src) -- (n1);
\draw[arr] (ew) -- (n2);
\draw[arr,dashed] (fe) -- (ir);
\draw[arr] (n1) -- node[below]{Btor2} (z3);
\draw[arr] (n2) -- node[below]{Btor2} (z3);
\end{tikzpicture}}
\caption{The \code{rtl-prove} check. The top row up to the header is the translator. The reference
(bottom left) runs no \code{svcc} code. Shaded dashed boxes are \code{svlang}, which all three
paths use and which the check trusts.}
\label{fig:gate}
\end{figure}
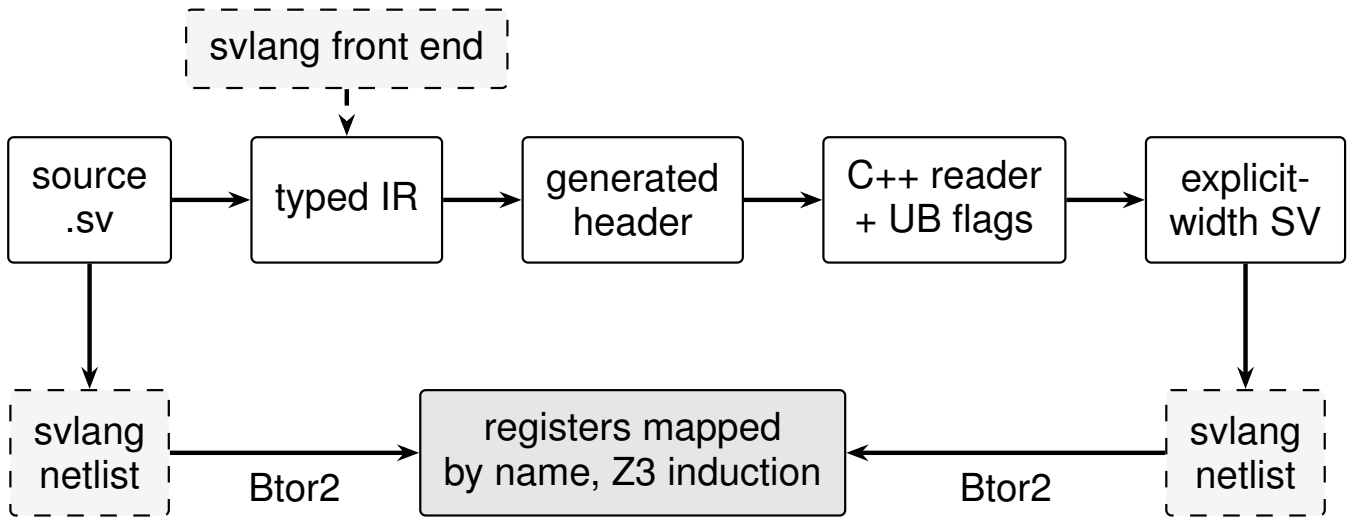

Fig.~\ref{fig:gate} shows the check. The reference side lowers the source text to a netlist with
\code{svlang} and writes Btor2~\cite{niemetz2018btor2}. No \code{svcc} code runs on that side. The
checked side starts from the generated header.

\subsection{A model of C++ evaluation}

The reader parses the subset of C++ that \code{rtl2sc} emits and gives each expression the value
C++17 and SystemC give it. \code{bool} and 8- and 16-bit integers promote to \code{int}.
\code{sc\_uint<W>} converts to \code{unsigned long long} and \code{sc\_int<W>} to \code{long long}
(the LP64 data model). Binary operators bring both operands to a common type by the usual
arithmetic conversions and compute in that type. A value is truncated only where the code
constructs or assigns a narrower object. \code{sc\_biguint} arithmetic is exact. A \code{write} in
an edge-triggered method is a nonblocking assignment. A level-sensitive method must be sensitive to
every signal it reads; it is read as combinational logic, or as a latch where it does not assign on
every path. Behavior the standard leaves to the implementation is fixed as C++20 defines it:
narrowing to a signed type wraps, and \code{{>}{>}} of a negative value is arithmetic.

\emph{Undefined behavior.} An operation whose result C++ or SystemC does not define gets no value.
It raises a 1-bit flag, an extra output of the lifted module that the proof must show stays at 0.
The flagged operations are signed overflow, division or modulo by zero, a shift by a negative
amount or by the width or more, a left shift of a negative value, and a select or index out of
range. Guards in the code (\code{?:}, \code{\&\&}, \code{if}) are kept, so a guard the generator
forgot produces a counterexample. I took this treatment from Alive2~\cite{lopes2021alive2}.

\emph{Explicit-width output.} The reader writes what it computed as SystemVerilog in which every
operator result is self-determined and every width change is a size cast, a concatenation or
\code{\$signed}. Context sizing in SystemVerilog then has nothing left to change. This is the step
that closes the gap of Section~\ref{sec:failure}. I chose to print SystemVerilog here, instead of
encoding C++ straight into Btor2, so that both sides of the proof go through the same netlist
lowering and so that I can read what the reader understood. Internally the reader builds the same
typed design IR that the converter uses and prints it with the same printer; Section~\ref{sec:assume}
comes back to this.

\subsection{Proof and verdicts}

Both texts are lowered by \code{svlang} to Btor2. Registers are mapped by hierarchical name and
width, one to one, and Z3~\cite{demoura2008z3} checks one inductive step: from any state in which
the mapped registers are equal, the outputs agree, the undefined-behavior flag is 0, and the mapped
registers are equal after a step. If that holds the verdict is \emph{proved}, for every run from
equal start states, with no bound. This is a register-mapped equivalence check, with no retiming
or state re-encoding. If the inductive step fails, a bounded check from reset (8 steps by default)
gives a concrete \emph{mismatch} or, if none is found, \emph{bounded}. A bounded verdict is printed
as \code{[BND]} and exits with code 2 unless \code{-{}-accept-bounded} is given. It is not counted
as a proof. A register without a partner, or no answer from Z3 on the inductive step, also leads to
the bounded check; no answer there is \emph{inconclusive} and counts as a failure. A design the
model does not cover (more than one clock, a multi-dimensional array port, or C++ the reader does
not model) gets \emph{not covered} and exit code 3. The check also proves the lowering (typed IR
against source) separately, which tells a lowering failure apart from an emitter failure.

The same machinery checks an untimed C++ reference function against pipelined RTL with
\code{prove-sec -{}-latency L}: from an arbitrary RTL state, with resets held inactive, the RTL
output equals the function of the inputs $L$ cycles earlier. This is C-to-RTL equivalence checking
as done in commercial tools~\cite{koelbl2009c2rtl}, and the same check of an untimed C++ algorithm
against sequential RTL that Olmos et al.\ use to start formal verification before the RTL is
friendly to it~\cite{olmos2024hlec}.

\subsection{What a pass assumes}
\label{sec:assume}

A pass holds only under the following assumptions.
(1)~Values are two-state: X and Z read as 0.
(2)~\code{svlang}'s elaboration is taken as the meaning of SystemVerilog. The translator, the
reference and the checked side all use it, so a front-end bug that changes the source's meaning is
invisible to the proof. Three such bugs were found by co-simulation (Section~\ref{sec:defects}),
and the proof could not have found them. The explicit-width text uses only a few plain constructs
to limit this exposure on the checked side.
(3)~There is one clock, and an asynchronous reset takes effect at the clock step.
(4)~The C++ reader is my model of C++ typing, not a compiler front end. I checked it against
SystemC~3.0.3 with a probe program and against compiled SystemC with the mutation experiment of
Section~\ref{sec:mutation}. The reader and the converter also share the typed IR and its printer,
so a printer bug could affect both.
(5)~Proved means equivalent from equal start states.
(6)~SystemC process scheduling is not modeled. A process made sensitive to the edge of an internal
copy of a clock is treated as if it were on the root clock. Such a header runs one delta cycle late
in a real kernel and still passes the proof; I confirmed this with a hand-edited header.
Co-simulation is the only check of scheduling.

\section{Evaluation}
\label{sec:eval}

I wanted answers to four questions. How much of a design suite does the check prove (RQ1)? Does
it reject wrong C++, judged by compiled SystemC and not by my own tool (RQ2)? How does the
conversion do on real designs, and what did those runs break (RQ3)? How does \code{svcc} compare
with a recent translator (RQ4)?

\emph{Setup.} \code{svlang}~0.5.1, Z3~5.4.1 (4.15.5 for the corpus proof), Verilator~5.052 and
SystemC~3.0.3 (my fork of the Accellera repository, commit \code{9767bff}, C++17). RQ1 was run at
\code{svcc} commit \code{00c80f0} on Linux x86\_64 and repeated on the same RTL code in a second
Linux container. The mutation experiment (RQ2) was run at commit \code{5dd247d} on macOS arm64. The
stress run (RQ3) was run at commit \code{c8e16fb} and its proof stage at commit \code{1db5bb2}, both
on Linux x86\_64. Every command and result is recorded in the repository's experiment log.

\subsection{RQ1: Proof coverage on the design suite}

The suite has \nDesigns{} designs: 24 construct-group designs in \code{testdata/rtl}, the
\nProbes{} probes of Section~\ref{sec:scav}, and \nCorpusRtl{} pipelined datapaths with C++ golden
models. I wrote all of them. The check proves \nProved{} by induction, for both the lowering and
the generated C++, and none of these passes is bounded-only. It reports \nNotCovered{} as not
covered: three have more than one clock or a computed clock (\code{clock\_gating},
\code{clock\_vector}, \code{edges\_misc}), one has a multi-dimensional array port
(\code{array\_ports}), and in one the reader does not model a local array copy
(\code{array\_elems}). None fails.

The designs are small. \nCombinational{} of the \nProved{} have no registers to map, so for them
the induction reduces to a combinational equivalence check; the largest has \nMaxPaired{} mapped
registers. Z3 takes 9 to 31~ms per design for the generated-C++ side. All \nGolden{} golden models
prove against their RTL with \code{prove-sec}.

One of those golden models caught me. I had written it for an earlier version of this work, and
it multiplied two \code{uint16\_t} values. In C++ both operands promote to \code{int}, the product
can overflow, and signed overflow is undefined. Testing with one compiler had never shown a
problem. The proof found it.

\subsection{RQ2: Does the check reject wrong C++?}
\label{sec:mutation}

\emph{Negative tests.} The two designs of Section~\ref{sec:failure} are regression tests that must
fail, together with a dropped sign conversion, an unguarded division, a missing sensitivity entry,
a wrong latency, a datapath bug in RTL, a golden model with promoted-multiply overflow, and C++
outside the subset. All fail.

\emph{Mutation experiment.} Negative tests that I chose can share my blind spots, so I compared
the proof with an oracle that shares no code with it: compiled SystemC simulated against Verilator
for 5,000 random steps. A script edited the generated header of \nMutDesignsGiven{} designs one
place at a time, changing the width or signedness at which C++ evaluates an expression.
\nMutDesigns{} designs produced mutants. Each mutant went to both judges (Table~\ref{tab:mut}).

\begin{table}[t]
\centering
\caption{Mutation experiment: proof against compiled SystemC co-simulated with Verilator, on mutants of \nMutDesigns{} designs. Edits: (a)~\code{sc\_uint<8>(e)} to \code{(e)}; (b)~to \code{sc\_uint<9>(e)}; (c)~to \code{sc\_int<8>(e)}; (d)~\code{uint64\_t(e)} to \code{(e)}. ``Proof accepts'' counts exit code 0.}
\label{tab:mut}
\footnotesize
\setlength{\tabcolsep}{2.5pt}
\begin{tabular}{@{}l r r r r r@{}}
\toprule
Edit & \shortstack{Com-\\piles} & \shortstack{Sim.\\differs} & \shortstack{Proof\\finds} & \shortstack{Reader\\rejects} & \shortstack{Proof\\accepts} \\
\midrule
drop truncation (a) & 133 & 7 & 4 & 3 & 0 \\
one bit wider (b) & 106 & 5 & 5 & 0 & 0 \\
flip signedness (c) & 28 & 5 & 5 & 0 & 0 \\
drop built-in cast (d) & 14 & 0 & 0 & 0 & 0 \\
\midrule
Total & \nMutCompiled{} & \nMutSimDiffers{} & \nMutProofFinds{} & \nMutReaderRejects{} & \textbf{\nMutAccepted{}} \\
\bottomrule
\end{tabular}
\end{table}

No mutant that the simulation showed to be wrong was accepted: \nMutProofFinds{} produced a
counterexample and \nMutReaderRejects{} were refused by the reader. Of the \nMutSimEqual{} mutants
that simulate equal, the proof accepted \nMutEqualAccepted{} and the reader refused
\nMutEqualRefused{}. The proof never reported a difference that the simulation did not see. Most
edits are harmless because an outer construction truncates again, so only \nMutSimDiffers{}
mutants change behavior, in 10 designs. The experiment was run at an earlier commit than RQ1 and
does not tell bounded from inductive passes apart.

All four edits touch casts, the class of error R1 was built for. I did not mutate sensitivity
lists, edge kinds, process sensitivity on internal signals, operators, constants or indices, and
assumption (6) says at least one of those would be accepted. A further \nMutNoCompile{} mutants did
not compile, and the reader accepted \nMutNoCompileRead{} of them, because it is not a type checker;
\code{rtl-prove} does not compile the header, co-simulation does. What this experiment shows is that
my C++ model agrees with real SystemC on cast and width edits in small designs.

\subsection{RQ3: Real designs}
\label{sec:stress}

Author-written designs only go so far, so I took every module of \nCodeBases{} open-source code
bases as a top with its default parameters: picorv32, SERV, biRISC-V, verilog-axi, SCR1,
tech\_cells\_generic, common\_cells, PULP AXI, CVFPU, Ibex, CV32E40P, CV32E40X, CVA6, Caliptra,
OpenTitan and BaseJump STL, \nModules{} modules in all. Each module passes through three stages:
convert, compile the header against SystemC, and co-simulate against Verilator for \nSteps{} random
steps, comparing every output after every step. Each step changes the data inputs and then toggles
one clock chosen at random, so a run has 1,500 clock cycles in all, about $1{,}500/k$ per clock for
a module with $k$ clocks. Both models run two-state, inputs that reach an edge sensitivity are
driven as clocks one at a time, synchronous resets are ordinary random inputs, and Verilator is
built without the \code{VERILATOR} macro so that both models come from the same text. I did not
measure coverage.

\nMatch{} modules convert, compile and match Verilator, and \nMismatch{} differ. The other
\nRejected{} are rejected by the front end or by \code{svcc} with a located error, or exceed 2~GB or
120~s. Many rejections are not real gaps: a module tried alone with placeholder defaults
(\code{parameter type T = logic}, a width of 0 or $-1$, an unconnected interface port) is not a
valid top, and the \code{svlang} front-end test corpus classifies \nPlaceholder{} such modules. The
largest genuine gaps are unpacked structs and struct-typed ports through type parameters,
interfaces, hierarchical references, and array ports connected to anything but a whole array.
Coverage is uneven across code bases: most of CVA6 and PULP AXI is rejected, and most matching
modules are leaf cells or blocks, not whole cores.

Converting clock-gating cells to enables (Section~\ref{sec:convert}) moved the result from
\nMatchBeforeCG{} to \nMatch{} matching modules and cut clock-related rejections from
\nClockRejBefore{} to \nClockRejAfter{}, with no module going from match to mismatch. None of the
\nMismatch{} remaining differences is an \code{svcc} defect. One is a Verilator defect: in a module
with a placeholder parameter, an out-of-range write lands on an in-range bit. The two in CVFPU's
cast unit, and the first difference in its fused multiply-add, come from a front-end width error on
\code{**} in a sized parameter, fixed by an unreleased \code{svlang} change (commit \code{af574c9}).
With that fix, the fused multiply-add differs only when the random format code names no format. The
module then reads its per-format arrays out of range, so the result is x under clause~11.5.1 of
IEEE~1800-2023. With the format inputs limited to valid codes, the two models agree for 200,000
steps.

\emph{Proof on the matching modules.} \code{rtl-prove} accepts multi-file designs, so I ran the
check on every module that matched Verilator, \nProofRecords{} records counting a module once per
code base. It proved \nCorpusProved{}, gave a bounded-only result for \nCorpusBounded{}, failed on
\nCorpusFail{} and reported \nCorpusNotCovered{} as outside its model, with no tool errors. Of the
proved modules, \nCorpusComb{} have no registers; the largest has \nCorpusMaxRegs{} mapped
registers, and Z3 took a median of 28~ms and at most 44~s per module. None of the failures is a
functional difference. In \nCorpusUnknown{} Z3 returned no answer, \nCorpusUnsupported{} use a
construct the reader rejects (eight are clock-gating latches written with a nonblocking
assignment), \nCorpusNoOutputs{} have no outputs to compare, and \nCorpusFalseDiff{} are false
differences caused by the check's start-state constraint on \code{integer} loop variables that the
front end infers as latches. Most modules outside the model have more than one clock or use
assertions with \code{\$isunknown}. Bounded-only results are not counted as proofs.

\emph{Defects found.}
\label{sec:defects}
Running one design through \code{svlang}, \code{svcc}, a SystemC kernel and Verilator and comparing
the results is a differential test of those tools, in the line of Csmith for C
compilers~\cite{yang2011csmith}, Verismith for synthesis tools~\cite{herklotz2020verismith}, and
ChiGen, which generates Verilog to test EDA tools and has found bugs in Verilator and
Yosys~\cite{vieira2025chigen}. I used real code bases instead of generated designs. Each defect
below gave a wrong result with no diagnostic.

\code{svcc}: (1)~A process holding a local copy of an array wrote every element back, including
elements driven by other processes, which lost updates or caused a multiple-driver abort. (2)~A
wildcard \code{.*} connection bound only its first port. This one bug accounted for 57 mismatching
Caliptra modules. (3)~A register clocked by a plain copy of the clock (\code{assign gclk = clk}) ran
one delta cycle late. Registers behind clock-gating cells did the same, which I already knew about,
and that led to the conversion of Section~\ref{sec:convert}. Each case now has a regression design.
Defect (3) is a scheduling effect, which only co-simulation checks (assumption (6)).

\code{svlang}: three elaboration defects, confirmed against IEEE~1800-2023 and fixed in
\code{svlang}~0.5.1. A generate \code{if} on a reduction of a parameter elaborated neither branch;
a fill literal \code{'1} as a sized parameter default evaluated to 1; and a select on a struct member
with its own range was bound against the wrong range. They change the netlist that every user of
that front end sees.

Verilator: (1)~A module that instantiates itself is built without its children when it is
selected with \code{-{}-top-module}, and its output stays 0. Here the reference was wrong and
\code{svcc} was right; \code{rtl-cosim} now wraps such a top. (2)~An out-of-range write through a
variable select changes an in-range bit when the vector has a negative bound, as in
\code{[-1:0]}, or when the index selects past a packed two-dimensional array whose size is not a
power of two. Clause~11.5.1 says such a write has no effect. Both are present in 5.020 and in the
development version (5.055, commit \code{a53cbcd}), and Icarus Verilog~12 gives the expected
result. I reported them as Verilator issues 9328 and 9329~\cite{verilator9328,verilator9329}.

\subsection{RQ4: Comparison with a recent translator}
\label{sec:scav}

Zheng et al.~\cite{zheng2026bidir} present a bidirectional translator between SystemC and Verilog.
I compared with SCAV, a public implementation of that translator (commit \code{2ef2e6b}). Its
Verilog-to-SystemC direction rewrites a text dump of a Verilog-2005 parse tree with regular
expressions, and its SystemC-to-Verilog direction recognizes the method names its own forward
direction generates. It does not check its output. I have not reproduced the article's own
evaluation.

I used two inputs: the \nCorpusRtl{} pipelined datapaths of RQ1, and \nProbes{} probes written for
this study, each targeting one construct (blocking assignments in combinational and sequential
blocks, signed arithmetic, parameters, wide vectors, memories, loops, \code{casez}, functions,
generate, part selects, a state machine, hierarchy, counters, the width rule, wrap-around, and
latch ordering). A result counts as correct if the output compiles against SystemC and matches the
source in co-simulation. SCAV's translator stopped with an internal error on 3 of the
\nCorpusRtl{} datapaths and 9 of the \nProbes{} probes. Its output compiled for 0 of the datapaths
and 2 of the probes, and was correct for none. Of the 9 examples shipped with SCAV, 3 compiled and 1
of those simulated wrongly. SCAV was judged against Icarus Verilog with SystemC 2.3.4.

\code{svcc} is correct on all \nCorpusRtl{} datapaths and all \nProbes{} probes, judged with
\code{rtl-cosim} against Verilator with SystemC 3.0.3, and the check proves them. So the two tools
ran under different harnesses. The probes are also part of \code{svcc}'s own suite and were used
during its development. The same probes run against \code{svcc}'s earlier text-based path gave 3
of \nProbes{}, and that is the result that exposed the failure of Section~\ref{sec:failure}. I read
this comparison as showing that SCAV does not handle these constructs. It is not a benchmark on
neutral ground.

\section{Related Work}
\label{sec:related}

\emph{HDL to SystemC and C translation.} HIFSuite~\cite{bombieri2010hifsuite} converts VHDL and
Verilog to SystemC and back through a common intermediate format, and Syed et
al.~\cite{syed2013vhdl2sc} translate RTL VHDL to SystemC with attention to process and delta-cycle
mapping. v2c~\cite{mukherjee2016v2c} translates synthesizable Verilog to C so that software
verifiers can check RTL, and FuzzWiz~\cite{gadde2024fuzzwiz} converts RTL to C/C++ models so that
software fuzzers can reach coverage faster than simulation. Zheng et al.~\cite{zheng2026bidir}
translate in both directions between SystemC and Verilog. Verilator~\cite{snyder2020verilator}
produces SystemC for simulation. Each of these trusts its C or SystemC to mean what the RTL means;
the check in this paper is one way to test that trust.

\emph{C and RTL consistency.} Clarke, Kroening and Yorav check a C program against Verilog by
unrolling both under their own semantics~\cite{clarke2003cverilog}, the approach behind hw-cbmc and
later C-to-RTL tools; CBMC~\cite{clarke2004cbmc} gives the bit-precise C semantics. Word-level
sequential equivalence between C/C++ models and RTL~\cite{koelbl2009c2rtl} is supported by
commercial tools, and Olmos et al.~\cite{olmos2024hlec} use it to check an untimed C++ model
against sequential RTL early in a project. R1 is the premise of all of this work. \code{rtl-prove}
applies it to the output of a translator, with a C++ model limited to the code that translator
emits, and \code{prove-sec} is a small open version of C-to-RTL checking. Yosys
EQY~\cite{yosys_eqy} checks equivalence between two RTL or netlist versions.

\emph{Translation validation and verified compilation.} Translation validation was proposed by
Pnueli et al.~\cite{pnueli1998tv}, applied to a production compiler by Necula~\cite{necula2000tv}
and to LLVM by Alive2~\cite{lopes2021alive2}. In hardware, Kundu et al.~\cite{kundu2010tvhls}
validate high-level synthesis with a simulation relation that need not be a one-to-one register
map, which is more general than my name-based mapping. Closest in spirit is the Verilog vectorizer
of Guimar\~aes et al.~\cite{guimaraes2026vec}. It lowers Verilog into CIRCT, rewrites groups of
bit-level assignments as word-level operations, and prints Verilog again, and the authors show each
of their 1,157 rewritten designs equivalent to its original with sequential equivalence checking in
Jasper. That is translation validation of a source-to-source HDL tool, per output, as here. Their
transformation stays inside one language, so the width rules cannot drift between input and
output; \code{svcc} crosses into C++, and that crossing is what R1 is about. The alternative to
validation is a verified translator: CompCert~\cite{leroy2009compcert} for C,
Vericert~\cite{herklotz2021vericert} for C to Verilog, and Lutsig~\cite{loow2021lutsig} for a
Verilog subset to netlists. These give stronger guarantees over smaller subsets.

\emph{Semantics.} M\"uller et al.~\cite{mueller2001scsem} give a formal simulation semantics of
SystemC, and Herber et al.~\cite{herber2008scta} model check SystemC designs, including the
scheduler, with timed automata. My proof does not model the scheduler at all.

\emph{Language models.} Language models generate Verilog~\cite{liu2023verilogeval} and translate
C++ to high-level-synthesis-ready code or Verilog~\cite{collini2024c2hlsc,lyu2026autoc2v}. Their
outputs are mostly judged by tests. A check that reads generated code with the target's semantics
could judge such outputs too, as long as they stay inside the subset the reader models. I have not
tried this yet.

\section{Limitations}
\label{sec:limits}

\begin{itemize}
\item Proofs cover single-clock designs with two-state values, with asynchronous reset applied at
the clock step. Multi-clock behavior, X, reset timing and SystemC process scheduling are checked
only by co-simulation. A header with the delta-late clock copy of defect (3) passes the proof.
\item \code{svlang} is trusted on both sides and is also the translator's front end. The C++ reader
is my own model of C++ and shares the typed IR and printer with the converter. I also wrote the
emitter, \code{svlang} and the mutation script.
\item The mutation experiment only edits casts.
\item On the open-source corpus, \nCorpusBounded{} modules have only a bounded result and Z3 gave no
answer on \nCorpusUnknown{}. The co-simulation has no coverage figure.
\item The SCAV comparison uses different harnesses for the two tools and probes that were part of
\code{svcc}'s development.
\item \code{svcc} also translates UVM testbench components, constraints, covergroups and assertions
to UVM-SystemC, CRAVE and FC4SC. Those checks still compare through \code{svcc}'s own
representation and have not been rebuilt to rule R1. This paper makes no claim about them.
\end{itemize}

\section{Conclusion and Next Steps}
\label{sec:conclusion}

My first check for a SystemVerilog-to-SystemC translator read the C++ output with SystemVerilog's
rules and passed wrong models. The fix was to read the generated C++ the way C++ reads it, to treat
undefined behavior as something to prove absent, and to build the reference without the
translator's code. On the designs I tried, that closed the gap. A mutation experiment judged by
compiled SystemC found no cast or width error that the check accepted, and on real code the check
proves \nCorpusProved{} of the \nMatch{} conversions that match Verilator. Scheduling is still left
to co-simulation.

There are two things I want to do next. The first is in the check itself: a scheduling check for
processes on internal clocks, multi-clock designs, and a separate verdict for solver unknowns. For
the \nCorpusUnknown{} corpus modules where Z3 gave no answer, a vectorization pass of the kind
Guimar\~aes et al.\ describe~\cite{guimaraes2026vec} is worth trying before the proof; they report
that it cuts Jasper's elaboration time by 28\% and its memory by 51\%. I have not tried this.

The second is the reason I care about checked translations at all. On large compute designs, such
as the ones built for datacenters, one RTL simulation regression can take hours, and an AI agent
that edits RTL needs a much faster answer before it tries the next change. I want to use
\code{svcc} to build a harness for such an agent. The agent would run its quick experiments on
SystemC models of the blocks it touches, and the proof would be the gate that says, in
milliseconds on blocks the size of those in this paper (the median Z3 time on the corpus was
28~ms), that each model still matches its RTL. I have not built or measured this loop, and whether
it holds on full datacenter-class designs is open.

\section*{Data and Code Availability}
\code{svcc} (Rust, \nRustLines{} lines, with Python bindings), the design suite, the probes, the
mutation and stress scripts, and all results are at \url{https://github.com/mayurkubavat/svcc}
under the Apache 2.0 license. Every release is archived on Zenodo~\cite{kubavat2026svcczenodo}.
Each measurement in this paper is recorded with its command, commit, machine and tool versions in
the repository's experiment log (\code{results/experiment\_log.jsonl}).

\section*{Acknowledgment}
The author thanks Fernando Magno Quint\~ao Pereira for direct comments on an earlier draft of this
paper, and the maintainers of Verilator, SystemC and Z3. The author used Google Antigravity IDE to
help develop the \code{svcc} implementation and to refine this manuscript, and other AI coding and
writing assistants to help prepare parts of the benchmark scripts, the evaluation runs and the
manuscript text. The author reviewed all of it, ran and checked every measurement, proof and
reference reported here, and takes full responsibility for the content.

\bibliographystyle{IEEEtran}
\bibliography{refs}
\end{document}